\documentclass[11pt]{article}
\usepackage{aas_macros,amsmath,amssymb,comment,cite,esint,graphicx,mathtools,diagbox}
\usepackage{bm}
\usepackage[margin=.8in,letterpaper]{geometry}
\usepackage[colorlinks=true]{hyperref}
\usepackage[affil-it]{authblk}
\usepackage{subcaption}
\usepackage[utf8]{inputenc}
\usepackage{mathrsfs}
\usepackage{appendix}
\usepackage{amssymb}
\usepackage{float}                  
\usepackage{color}
\usepackage{cite}
\usepackage{hyperref}
\hypersetup{pageanchor=false}
\usepackage{indentfirst}
\usepackage{url}
\usepackage{xfrac}
\usepackage{caption}
\usepackage[numbers,square,comma,sort&compress,merge]{natbib}
\usepackage{esint}
\usepackage{overpic}
\usepackage{graphicx}
\usepackage{epsf,amsmath,bbold,amsfonts,stmaryrd}
\usepackage{textcomp}
\usepackage{ulem}
\usepackage{tikz}
\usepackage{multirow}
\numberwithin{equation}{section}
\usepackage{amsfonts}

\newcommand{\be}{\begin{equation}}
\newcommand{\ee}{\end{equation}}
\newcommand{\bea}{\setlength\arraycolsep{2pt} \begin{eqnarray}}
\newcommand{\eea}{\end{eqnarray}}
\newcommand{\nn}{\nonumber}
\newcommand{\mm}{\mathrm}
\newcommand{\mc}{\mathcal}

\def\fft#1#2{{\frac{#1}{#2}}}

\def\0{{\sst{(0)}}}
\def\1{{\sst{(1)}}}
\def\2{{\sst{(2)}}}
\def\3{{\sst{(3)}}}
\def\4{{\sst{(4)}}}
\def\5{{\sst{(5)}}}
\def\6{{\sst{(6)}}}
\def\7{{\sst{(7)}}}
\def\8{{\sst{(8)}}}
\def\sst#1{{\scriptscriptstyle #1}}

\begin{document}
\title{\bf Magnetic reconnection under centrifugal and gravitational electromotive forces}
\author{
Zhong-Ying Fan$^{1}$, Fan Zhou$^{2, 3}$, Yuehang Li$^{2, 3}$,  Minyong Guo$^{2,3\ast}$, Bin Chen$^{4,5,6,7}$}
\date{}

\maketitle

\vspace{-10mm}

\begin{center}
{\it
$^1$ Department of Astrophysics, School of Physics and Electronic Engineering, Guangzhou University, Guangzhou 510006, P. R. China\\\vspace{4mm}

$^2$ School of physics and astronomy, Beijing Normal University,
Beijing 100875, P. R. China\\\vspace{4mm}

$^3$Key Laboratory of Multiscale Spin Physics (Ministry of Education), Beijing Normal University, Beijing 100875, China\\\vspace{4mm}

$^4$ Institute of Fundamental Physics and Quantum Technology, Ningbo University, Ningbo, Zhejiang 315211, China\\\vspace{4mm}

$^5$ School of Physical Science and Technology, Ningbo University, Ningbo, Zhejiang 315211, China\\\vspace{4mm}

$^6$Department of Physics, Peking University, No.5 Yiheyuan Rd, Beijing
100871, P.R. China\\\vspace{4mm}

$^7$Center for High Energy Physics, Peking University,
No.5 Yiheyuan Rd, Beijing 100871, P. R. China\\\vspace{2mm}



}
\end{center}

\vspace{8mm}

\begin{abstract}

We examine the physical implications of the centrifugal and gravitational electromotive forces on magnetic reconnection in a Kerr black hole background. We find that both forces increase the reconnection rate, though the underlying mechanisms differ substantially. The gravitational force leads to a separation of charge density, breaking the quasi-neutrality of the plasma. In contrast, the centrifugal electromotive force affects the electric current by reducing the effective length of the current sheet. This reduction arises from the non-Euclidean spatial geometry observed by a locally comoving observer with respect to the rotating sheet. This phenomenon amplifies both the transport of charged carriers and the thermal-inertia effect within the current sheet, irrespective of the presence of a black hole.
\end{abstract}

\vfill{\footnotesize $\ast$ Corresponding author: minyongguo@bnu.edu.cn;}

\thispagestyle{empty}

\pagebreak

\tableofcontents
\addtocontents{toc}{\protect\setcounter{tocdepth}{2}}




\section{Introduction}

Magnetic reconnection can rapidly convert magnetic energy into particle energy for magnetohydrodynamical plasmas. It is usually considered to be a primary candidate to explain various high energy emissions \cite{book,biskamp}, such as magnetospheric substorms, stellar flares and jets from gamma-ray bursts and from active galactic nuclei.
In recent years the importance of relativistic effect for strong magnetized plasmas was recognized \cite{Zenitani:2001fef,Lyubarsky:2008yi,Lyubarsky:2005zt}. In cases where magnetic energy largely exceeds the particle kinetic energy, the outflow velocity will be mildly relativistic. The generalizations of magnetic reconnection to relativistic regimes have been widely conducted in the literature
\cite{Zenitani:2009bj,Liu:2014ada,Zenitani:2010vq,Takahashi:2011br,Kowal:2012rv,Kowal:2011yz,Zenitani:2007vt,Cerutti:2011cv}. It was established that relativistic effect is of paramount importance for highly efficient conversion of magnetic energy and particle acceleration. Various studies suggest that the relativistic magnetic reconnection is a competitive mechanism for nonthermal emissions in various strongly magnetically dominated environments in astrophysics \cite{Guo:2014via,Sironi:2014jfa,Guo:2015cua}.

Recently, the effects of gravitation on magnetic reconnection has been partly explored in the literature. In \cite{Asenjo:2017gsv}, the Sweet-Parker model was firstly generalised to curved spacetime. It was established that the spacetime curvature supresses the reconnection rate in comparison to the flat spacetime limit. The applications of magnetic reconnection to energy extraction from rotating black holes have been carried out in \cite{Comisso:2020ykg,Chen:2024ggq,Shen:2024sdr,Shen:2024plw,Camilloni:2024tny}. In \cite{Fan:2024fcy}, a Petschek like model was developed for a rotating reconnection layer in a Kerr black hole background. In \cite{Comisso:2014nva}, it was shown that for relativistic plasmas that retain thermal-inertia effect, the reconnection rate was increased for both the Sweet-Parker and the Petschek configuration. The work was generalised to curved spacetime in \cite{Comisso:2018ark}. The effect of gravitational electromotive force on a nonrotating configuration in a Schwarzschild black hole background was examined in \cite{Asenjo:2019nji}.

Motivated by recent advancements, this study aims to extend the analysis presented in \cite{Asenjo:2019nji} by investigating a rotating reconnection layer within the context of a Kerr black hole. Given that the reconnection layer typically does not corotate with the black hole, it is imperative to examine magnetic reconnection within a locally corotating frame (LCRF) with respect to the rotating layer. This approach enables a comprehensive exploration of the influences exerted by both centrifugal and gravitational electromotive forces on the process of magnetic reconnection. We will demonstrate that the effect of the black hole's spin is negligible compared to the gravitational influence of the black hole's mass. However, additional rotation of the reconnection layer, even of the same order as the black hole's spin, can significantly enhance the reconnection rate and the gravitational electromotive force. The underlying mechanisms in these two cases, however, are substantially different. The gravitational force leads to the separation of the charge density, breaking the quasi-neutrality of the plasmas \cite{Asenjo:2019nji}, whereas the centrifugal force acts on the electric current, causing the electromotive force, primarily due to the non-Euclidean spatial geometry for a comoving observer. The latter decreases the effective length of the current sheet, thereby enhancing both the charged carrier transport and the thermal-inertia effect, irrespective of the existence of a black hole.

The paper is organized as follows. In section 2, we set up the magnetohydrodynamical plasmas model which retains the thermal-inertia effect. Then we introduce a generally rotating disk in a Kerr black hole background and define the locally corotating frame. In section 3, we derive all dynamical equations for the plasmas using physical quantities in the locally corotating frame. In section 4, we examine various important properties for the rotating reconnection layer in a Kerr black hole background. In particular, we extract the leading contributions of the centrifugal and the gravitational electromotive force to the reconnection rate. We briefly conclude in section 5.

\section{Rotating MHD fluids}
Consider relativistic magnetohydrodynamic (MHD) plasmas in a curved spacetime. Here we would like to explore collisionless effects in the magnetic reconnection. The energy-momentum equation that retains thermal-inertia effects is \cite{Koide:2009dt}
\be \nabla_\nu\left[ h \Big(U^\mu U^\nu+\fft{\xi}{4n^2e^2}\,J^\mu J^\nu \Big) \right]=-\nabla^\mu p+J^\nu F^\mu_{\,\,\,\nu} \,,\label{emcon}\ee
where $n$ is the total number density of particles, $e$ is the electron charge, $h$ is the enthalpy density, $p$ is the proper plasma pressure, $U^\mu$ is the fluid four velocity and $J^\mu$ is the electric current. The parameter $\xi\approx 0$ for an electron-ion plasma and $\xi=1$ for a pair plasma, respectively. The Ohm's law is generalised to
\be\label{generalisedohm} U^\nu F^\mu_{\,\,\,\nu}=\eta(J^\mu-\rho_e U^\mu)+\fft{\xi}{4e^2n}\nabla_\nu\left[\fft{h}{n}\Big( U^\mu J^\nu+J^\mu U^\nu-\fft{\mu}{ne}J^\mu J^\nu \Big) \right] \,,\ee
where $\eta$ is the electrical resistivity, $\rho_e=-U_\mu J^\mu$ is the invariant charge density measured in the local center-of-mass frame and $\mu=(m_+-m_-)/(m_++m_-)$, where $m_\pm$ indicating the mass of corresponding charged particle.

Besides, MHD fluids obey 
the continuity equation \cite{Lichnerowicz}
\be \nabla_\mu(n U^\mu)=0 \,,\ee
 and the Maxwell's equations
\be \nabla_\nu F^{\mu\nu}=J^\mu \,,\qquad \nabla_\nu {}^*F^{\mu\nu}=0 \,.\ee

To study the magnetic reconnection for a rotating reconnection layer, we consider the corotating configuration in a Kerr black hole, whose metric can be written in the form of
\be ds^2=-\alpha^2 dt^2+\sum_{i=1}^3\big( h_idx^i-\alpha \beta^i  dt\big)^2 \,,\ee
where $h_0^2=-g_{00}\,,h_i^2=g_{ii}$. Here and below, for all equations expressed in the 3+1 formalism, we will write down the summation over the repeated indices explicitly or no summation will be implied for the repeated indices. The parameter $\beta^i$ is related to the black hole rotation $\beta^i=h_i\omega_i/\alpha\,,\omega_i=-g_{0i}/h_i^2$ and $\alpha^2=h_0^2+\sum_i (h_i\omega_i)^2$. An advantage of this coordinate is one can introduce the zero-angular-momentum-observer (ZAMO) frame $\{\hat t\,,\hat x^i\}$ \cite{Bardeen:1972fi}
\be d\hat{t}=\alpha dt\,,\qquad d\hat{x}^i=h_idx^i-\alpha\beta^i dt \,,\label{zamo}\ee
which corotates with the black hole. Previously various studies for magnetic reconnection have been conducted in this frame \cite{Asenjo:2017gsv,Comisso:2020ykg,Shen:2024sdr,Chen:2024ggq,Fan:2024fcy,Comisso:2018ark,Asenjo:2019nji}.
However, in this work, we will relax the condition and explore the magnetic reconnection for a generally rotating reconnection layer.

Before that, let us first present the metric components for a Kerr black hole in the Boyer-Lindquist (BL) coordinates $(x^0\,,x^1\,,x^2\,,x^3)=(t\,,r\,,\theta\,,\phi)$ 
\bea
&& h_0=\sqrt{1-\fft{2r_g r}{\Sigma^2}}\,,\qquad h_1=\fft{\Sigma}{\sqrt{\Delta}}\,,\nn\\
&&h_2=\Sigma\,,\qquad h_3=\fft{\Pi\sin\theta}{\Sigma}\,,
\eea
where $r_g=GM$ is the gravitational radius and $a=J/J_{\mm{max}}\leq 1$ is the rotation parameter ($J$ is the angular momentum and $J_{\mm{max}}=GM^2$). The functions $\Sigma\,,\Delta$ and $\Pi$ are given by
\bea
&&\Sigma^2=r^2+a^2r_g^2\cos^2\theta\,,\nn\\
&&\Delta=r^2-2r_gr+a^2r_g^2\,,\nn\\
&&\Pi^2=\big(r^2+a^2r_g^2 \big)^2-\Delta\big(ar_g\sin\theta \big)^2\,.
\eea
In addition, the black hole rotates only around the $\phi$-direction 
$\omega_1=\omega_2=0$ and $\omega_3=2r_g^2 ar/\Pi^2$ so that $\beta^i=\beta \delta^{i\phi}$. In the asymptotically weak gravity region, $\beta$ behaves as $\beta\sim 2ar_g^2/r^2$ to leading order at large r.

\subsection{General rotating layer}
In the Kerr black hole background, we consider a generally rotating disk
\be d\bar{t}=\alpha dt\,,\quad d\bar{x}^i=h_i dx^i-\alpha \bar{\beta}^i dt \,,\ee
where in general $\bar{\beta}^i\neq \beta^i$ and hence this frame does not corotate with the black hole. We let $\bar{\beta}^i-\beta^i=w^i$ and $w^i=w \delta^{i\phi}$. It turns out that the metric in this frame has mixing terms $\bar{g}_{0i}\neq 0$. One has
\be ds^2=-d\bar{t}^2+\sum_i(d\bar{x}^i+w^i d\bar{t}\,)^2 \,.\ee
We are interested in examining the magnetic reconnection in a locally corotating frame $\{\tilde{t}\,,\tilde{x}^i\}$, which is defined as
\be\label{crf} 
d\tilde{t}=q\Big( d\bar{t}-\sum_i q^{-2}w^i d\bar{x}^i \Big) \,,\quad d\tilde{x}^i=q_i^{-1}d\bar{x}^i\,,
\ee
where $q_i=\sqrt{1-(w^i)^2}$ and $q=\sqrt{1-w^2}$.
It is straightforward to see that in the LCRF, the metric becomes hypersurface orthogonal and is locally flat $ds^2=\eta_{\mu\nu}d\tilde{x}^\mu d\tilde{x}^\nu$. This is equivalent to building an orthonormal frame for the comoving observer. Though this is very convenient for us in the remaining of this work, it hides an important fact: the spatial geometry for a comoving observer becomes non-Euclidean. Since this is crucial for our physical explanations at the end of this work, we shall clarify it further. In standard approach, an ordinary comoving observer for a rotating disk is generally defined as a time-like curve $\{ \breve{t}\,,\breve{x}^i \}$ 
\be\label{crftrue} 
d\breve{t}=d\bar{t}-\sum_i q^{-2}w^i d\bar{x}^i \,,\quad d\breve{x}^i=d\bar{x}^i\,.
\ee
The metric for the comoving observer becomes
\be\label{noneuclidean} ds^2=-q^2 d\breve{t}^2+ \big(d\breve{x}^1 \big)^2+\big(  d\breve{x}^2 \big)^2+q^{-2}\big(  d\breve{x}^3\big) \,.\ee
Clearly the spatial slice defined at constant $\breve{t}$ becomes non-Euclidean. While the warped factor in the spatial geometry has been absorbed in our definition of the LCRF, its physical effect will be retained in the magnetic reconnection process. We will return to this point at the end of section 4.

To proceed, we present the transformation between the LCRF and the BL coordinates
\bea\label{inversecrf}
&& dt=\big(\alpha q \big)^{-1}\Big( d\tilde{t}+\sum_i w^i d\tilde{x}^i \Big) \,,\nn\\
&& dx^i=\big(q_ih_i \big)^{-1}\Big(\bar{\beta}^i d\tilde t+(1+w^i\beta^i)d\tilde{x}^i \Big)\,.
\eea
Clearly unlike the ZAMO frame (\ref{zamo}), the LCRF is strongly influenced by extra rotation of the reconnection layer (that is $w^i\neq 0$). Since we will examine various important properties of magnetic reconnection in the LCRF, here we would like to first translate vectors or tensors in the BL coordinates into those in the LCRF. According to (\ref{inversecrf}), one has for a vector $H^\mu$
\bea
&& H^0=\big( \alpha q\big)^{-1}\Big( \tilde{H}^0+\sum_j w^j\tilde{H}^j \Big)\,,\nn\\
&&H^i=\big( q_i h_i \big)^{-1}\Big( \bar{\beta}^i\tilde{H}^0+(1+w^i\beta^i)\tilde{H}^i \Big)\,,
\eea
and for a symmetric tensor $S^{\mu\nu}=S^{\nu\mu}$
\be
S^{00}=\big( \alpha q\big)^{-2} \tilde{\mathcal{S}}^{00}\,,\quad S^{0i}=\big( \alpha q q_i h_i\big)^{-1}\tilde{\mathcal{S}}^{0i}\,, \quad S^{ij}=\big( q_iq_jh_ih_j\big)^{-1}\tilde{\mathcal{S}}^{ij}\,,
\ee
 where
\bea
&&\tilde{\mathcal{S}}^{00}=\tilde{S}^{00}+2\sum_k w^k \tilde{S}^{0k}+\sum_{kl}w^k w^l\tilde{S}^{kl}    \,,\nn\\
&&\tilde{\mathcal{S}}^{0i}=\big(1+(w^i)^2+2w^i\beta^i\big)\tilde{S}^{0i}+\bar{\beta}^i\tilde{S}^{00}+\sum_k (1+w^i\beta^i) w^k\tilde{S}^{ki}      \,,\nn\\
&&\tilde{\mathcal{S}}^{ij}=(1+w^i\beta^i)(1+w^j\beta^j)\tilde{S}^{ij}+\bar{\beta}^i(1+w^j\beta^j)\tilde{S}^{j0}+\bar{\beta}^j(1+w^i\beta^i)\tilde{S}^{i0}+\bar{\beta}^i\bar{\beta}^j\tilde{S}^{00}      \,.
\eea
It turns out that in a Kerr black hole background, any equation $\nabla_\nu S^{\mu\nu}=H^\mu $ can be expressed in terms of the LCRF quantities as
\bea\label{spatialgeneral}
&& \fft{\partial }{\partial t}\big(\tilde{\mathcal{S}}^{0i}-\beta^i\tilde{\mc{S}^{00}} \big)+\fft{ q q_i}{h_1h_2h_3}\sum_{j}\fft{\partial}{\partial x^j}\left( \fft{\alpha h_1h_2h_3}{q_iq_jh_j}\,\big(\tilde{\mathcal{S}}^{ij}-\beta^i\tilde{\mc{S}}^{0j} \big) \right) \nn\\
&&+\fft{\tilde{\mathcal{S}}^{00}}{qh_i}\fft{\partial\alpha}{\partial x^i} 
-\alpha q q_i^2\sum_{j}\left[G_{ij}\big(\tilde{\mathcal{S}}^{ij}-\beta^i \tilde{\mc{S}}^{0j} \big)-G_{ji}\big(\tilde{\mathcal{S}}^{jj}-\beta^j\tilde{\mc{S}}^{0j} \big)\right] \nn\\
&&+\sum_j \sigma_{ji}\big( \tilde{\mc{S}}^{0j}-\beta^j\tilde{\mc{S}}^{00}\big)=\alpha q \Big((1+w^i\beta^i)\tilde{H}^i+w^i\tilde{H}^0 -\beta^i\sum_k w^k\tilde{H}^k\Big)\,,
\eea
where $G_{ij}=-(1/q_i^2q_jh_ih_j)\partial h_i/\partial x^j$ and $\sigma_{ij}=1/q_ih_j\partial(\alpha\beta^i)/\partial x^j$. The temporal component reads 
\bea\label{timegeneral}
&& \fft{\partial \tilde{\mathcal{S}}^{00}}{\partial t}+\fft{ q^2}{h_1h_2h_3}\sum_{j=1}^3\fft{\partial}{\partial x^j}\left( \fft{\alpha h_1h_2h_3}{qq_jh_j}\,\tilde{\mathcal{S}}^{0j} \right)+\sum_j\fft{q}{h_j}\fft{\partial\alpha}{\partial x^j}\big(\tilde{\mathcal{S}}^{0j}-\beta^j\tilde{\mc{S}}^{00} \big) \nn\\
&& +\sum_{jk}\alpha q^3 \beta^k\left[ G_{kj} \big( \tilde{\mc{S}}^{kj}-\beta^k\tilde{\mc{S}}^{0j}-\beta^j\tilde{\mc{S}}^{0k}+\beta^k\beta^j\tilde{\mc{S}}^{00} \big)-G_{jk}\big( \tilde{\mc{S}}^{jj}-2\beta^j\tilde{\mc{S}}^{0j}+(\beta^j)^2\tilde{\mc{S}}^{00} \big) \right]  \nn\\
&&+\sum_{jk}q^2\sigma_{jk}\big( \tilde{\mc{S}}^{jk}-\beta^j\tilde{\mc{S}}^{0k}-\beta^k\tilde{\mc{S}}^{0j}+\beta^j\beta^k\tilde{\mc{S}}^{00} \big)
=\alpha q \big( \tilde{H}^0+\sum_k w^k \tilde{H}^k  \big)\,.
\eea
These relations are valid to both the energy-momentum equation and the generalised Ohm's law. In the $w\rightarrow 0$ limit, the results reduce to those in \cite{Koide:2009dt} (see appendix A in that paper).  In addition, an antisymmetric tensor, for example the field strength tensor of Maxwell fields, is transformed as
\bea\label{ftensor}
&& F^{0i}=\big(\alpha q q_i h_i \big)^{-1}\Big[\big(1+w^i\beta^i \big) \big(\tilde{F}^{0i}+\sum_k w^k\tilde{F}^{ki}\big)+\bar{\beta}^i\sum_k w^k\tilde{F}^{k0} \Big]\,,\nn\\
&& F^{ij}=\big( q_i q_j h_i h_j\big)^{-1}\Big[ \big(1+w^i\beta^i \big)\big(1+w^j\beta^j \big)\tilde{F}^{ij}+\big(1+w^j\beta^j\big)\bar{\beta}^i\tilde{F}^{0j}-\big(1+w^i\beta^i \big)\bar{\beta}^j\tilde{F}^{0i} \Big]\,.
\eea
Application of these relations to dynamical equations of MHD will be given in the next section.

\section{Dynamical equations in LCRF}
Consider the continuity equation at first. In the LCRF, it can be rewritten as
\be \fft{\partial}{\partial t}\Big( n\tilde{\gamma}(1+\mathbf{w}\cdot \tilde{\mathbf{v}})\Big)+\fft{ q}{h_1h_2h_3}\sum_i\fft{\partial}{\partial x^i}\left[\fft{\alpha h_1h_2h_3}{q_ih_i}\,n\tilde{\gamma}\Big((1+w^i\beta^i)\tilde{v}^i+\bar{\beta}^i\Big) \right]=0 \,, \label{flux}\ee
where $\tilde{v}^i$ is the velocity of plasmas in the LCRF and $\tilde{\gamma}=(1-\tilde{\mathbf{v}}^2)^{-1/2}$ is the Lorentz factor. Here and below, the boldfaced letters stand for spatial vectors.

According to the Maxwell's equation,
\be J^\nu F^{\mu}_{\,\,\,\nu}=-\nabla_\nu T^{\mu\nu}_{\mm{em}}\,,\quad T^{\mu\nu}_{\mm{em}}=F^{\mu\sigma}F^\nu_{\,\,\,\sigma}-\fft14 g_{\mu\nu}F^2 \,.\ee
The energy-momentum equation (\ref{emcon}) can be written compactly as
\be \nabla_\nu T^{\mu\nu}=0\,,\qquad T^{\mu\nu}= p g^{\mu\nu}+h\Big(U^\mu U^\nu+\fft{\xi}{4n^2e^2}\,J^\mu J^\nu \Big)+T^{\mu\nu}_{\mm{em}} \,.\ee
We are mainly interested in the spatial components of the momentum equation. Using (\ref{spatialgeneral}) , we deduce
\bea\label{momenta}
&& \fft{\partial }{\partial t}\Big(\tilde{\mathcal{P}}^i-\beta^i(\varepsilon+\tilde\gamma\rho) \Big)+\fft{q q_i}{h_1h_2h_3}\sum_{j=1}^3\fft{\partial}{\partial x^j}\left( \fft{\alpha h_1h_2h_3}{q_iq_jh_j}\,\Big(\tilde{\mathcal{T}}^{ij}-\beta^i\tilde{\mc{P}}^j \Big)\right)+\fft{ \varepsilon+\tilde{\gamma}\rho }{qh_i}\fft{\partial\alpha}{\partial x^i} \nn\\
&&
-\alpha q q_i^2\sum_{j}\left[G_{ij}\big( \tilde{\mathcal{T}}^{ij}-\beta^i \tilde{\mc{P}}^j \big)-G_{ji}\big(\tilde{\mathcal{T}}^{jj} -\beta^j \tilde{\mc{P}}^j \big) \right]+\sum_j \sigma_{ji}\Big(\tilde{\mathcal{P}}^j-\beta^j(\varepsilon+\tilde\gamma\rho) \Big)=0\,,
\eea
where 
\bea
&&\tilde{\mc{P}}^i=\big[ 1+(w^i)^2+2w^i\beta^i\big] \tilde{P}^i+(\varepsilon_0+\tilde{\gamma}\rho)\bar{\beta}^i+(1+w^i\beta^i)\sum_k w^k \tilde{T}^{ki}\,,\\
&& \tilde{P}^i=h\tilde{\gamma}^2\tilde{v}^i+\fft{h\xi}{4n^2e^2}\,\tilde{J}^0\tilde{J}^i+\sum_{jk}\epsilon^{ijk}\tilde{E}_j \tilde{B}_k\,,
\eea
and
\bea
&&\tilde{\mathcal{T}}^{ij}=(1+w^i\beta^i)(1+w^j\beta^j)\tilde{T}^{ij}+(1+w^j\beta^j)\bar{\beta}^i\tilde{P}^{j}+(1+w^i\beta^i)\bar{\beta}^j \tilde{P}^{i}+(\varepsilon_0+\tilde{\gamma}\rho)\bar{\beta}^i \bar{\beta}^j\,,\\
&& \tilde{T}^{ij}=p\delta^{ij}+h\tilde{\gamma}^2\tilde{v}^i\tilde{v}^j+\fft{h\xi}{4n^2e^2}\,\tilde{J}^i\tilde{J}^j+\fft12\big(\tilde{E}^2+\tilde{B}^2\big)\delta^{ij}-\tilde{E}^i \tilde{E}^j-\tilde{B}^i \tilde{B}^j\,.
\label{stresstensor}\eea
Here $\varepsilon_0$ stands for the energy density measured in the LCRF and 
\bea
&& \varepsilon=\varepsilon_0+2\sum_i w^i \tilde{P}^i+\sum_{ij}w^iw^j \tilde{T}^{ij} \,,\\
&&\varepsilon_0=h\tilde{\gamma}^2+\fft{h\xi}{4n^2e^2}\big(\tilde{J}^0\big)^2-p-\tilde{\gamma}\rho+\fft12\big( \tilde{E}^2+\tilde{B}^2\big)\,.
\eea
Notice that $\tilde{J}^0$ is the separation of charge density measured in the LCRF. It is related to the invariant charge density as $\rho_e=-\tilde{U}_\mu \tilde{J}^\mu$. Previously it was shown \cite{Asenjo:2019nji} that a nonzero $\tilde{J}^0$ will affect the magnetic reconnection dramatically by the gravitational electromotive force. As a comparison, we will show that because of the non-Euclidean spatial geometry, the charge current density $\tilde{J}^i$ will affect the magnetic reconnection process significantly as well. This effect is independent of the gravitational force and can exist in the flat spacetime limit.

Similarly, using (\ref{spatialgeneral}), the generalised Ohm's law in the LCRF reads
\bea\label{spatialohm}
&& \fft{1}{ne}\fft{\partial }{\partial t}\big(\tilde{\mathcal{K}}^{0i}-\beta^i\tilde{\mc{K}^{00}} \big)+\fft{ q q_i}{ne h_1h_2h_3}\sum_{j}\fft{\partial}{\partial x^j}\left( \fft{\alpha h_1h_2h_3}{q_iq_jh_j}\,\big(\tilde{\mathcal{K}}^{ij}-\beta^i\tilde{\mc{K}}^{0j} \big) \right) \nn\\
&&+\fft{1}{ne}\fft{\tilde{\mathcal{K}}^{00}}{qh_i}\fft{\partial\alpha}{\partial x^i} 
-\fft{\alpha q q_i^2}{ne}\sum_{j}\left[G_{ij}\big( \tilde{\mathcal{K}}^{ij}-\beta^i \tilde{\mc{K}}^{0j}\big)-G_{ji}\big(\tilde{\mathcal{K}}^{jj}-\beta^j\tilde{\mc{K}}^{0j} \big)\right] \nn\\
&&+\fft{1}{ne}\sum_j \sigma_{ji}\big( \tilde{\mc{K}}^{0j}-\beta^j\tilde{\mc{K}}^{00}\big)=\alpha q \Big((1+w^i\beta^i)\tilde{H}^i +w^i\tilde{H}^0-\beta^i\sum_k w^k\tilde{H}^k\Big)\,,
\eea
where 
\bea
&&\tilde{\mathcal{K}}^{00}=\tilde{K}^{00}+2\sum_k w^k \tilde{K}^{0k}+\sum_{kl}w^k w^l\tilde{K}^{kl}    \,,\nn\\
&&\tilde{\mathcal{K}}^{0i}=\big(1+(w^i)^2+2w^i\beta^i\big)\tilde{K}^{0i}+\bar{\beta}^i\tilde{K}^{00}+\sum_k (1+w^i\beta^i) w^k\tilde{K}^{ki}      \,,\nn\\
&&\tilde{\mathcal{K}}^{ij}=(1+w^i\beta^i)(1+w^j\beta^j)\tilde{K}^{ij}+\bar{\beta}^i(1+w^j\beta^j)\tilde{K}^{j0}+\bar{\beta}^j(1+w^i\beta^i)\tilde{K}^{i0}+\bar{\beta}^i\bar{\beta}^j\tilde{K}^{00}      \,,
\eea
and 
\bea 
&&\tilde{K}^{00}=\fft{h\xi\tilde{\gamma}}{2ne}\,\tilde{J}^0\,,\nn\\  
&&\tilde{K}^{0i}=\fft{h\xi\tilde{\gamma}}{4ne}\,\big(\tilde{J}^i+\tilde{J}^0 \tilde{v}^i\big)\,,\nn\\
&& \tilde{K}^{ij}=\fft{h\xi\tilde{\gamma}}{4ne}\,\big( \tilde{v}^i\tilde{J}^j+\tilde{v}^j\tilde{J}^i  \big)\,,\nn\\
&& \tilde{H}^0=\tilde{\gamma}\tilde{v}^i\tilde{E}_i+\eta\tilde{\gamma}^2\tilde{v}^2\tilde{J}^0\,,\nn\\
&& \tilde{H}^i=\tilde{\gamma} \big( \tilde{E}^i+\sum_{jk} \epsilon^{ijk}\tilde{v}_j \tilde{B}_k \big)-\eta(\tilde{J}^i-\rho_e\tilde{\gamma}\tilde{v}^i )\,.
\eea
Notice that we have dropped the quadratic term of $\tilde{J}^\mu$, which is negligible in the weak gravity region, compared to the linear term in (\ref{generalisedohm}). The contribution of the various terms in (\ref{spatialohm}) to the electromotive force will be discussed later (see (\ref{ergeneral}) and below).

The temporal component of the generalised Ohm's law is given by
\bea\label{timeohm}
&& \fft{1}{ne}\fft{\partial \tilde{\mathcal{K}}^{00}}{\partial t}+\fft{ q^2}{neh_1h_2h_3}\sum_{j}\fft{\partial}{\partial x^j}\left( \fft{\alpha h_1h_2h_3}{qq_jh_j}\,\tilde{\mathcal{K}}^{0j} \right)+\fft{q}{ne}\sum_j\fft{1}{h_j}\fft{\partial\alpha}{\partial x^j}\big(\tilde{\mathcal{K}}^{0j}-\beta^j\tilde{\mc{K}}^{00} \big) \nn\\
&& +\fft{\alpha q^3}{ne}\sum_{jk} \beta^k\left[  G_{kj} \big( \tilde{\mc{K}}^{kj}-\beta^k\tilde{\mc{K}}^{0j}-\beta^j\tilde{\mc{K}}^{0k}+\beta^k\beta^j\tilde{\mc{K}}^{00} \big)-G_{jk}\big( \tilde{\mc{K}}^{jj}-2\beta^j\tilde{\mc{K}}^{0j}+(\beta^j)^2\tilde{\mc{K}}^{00} \big) \right] \nn\\
&&+\fft{q^2}{ne}\sum_{jk}\sigma_{jk}\big( \tilde{\mc{K}}^{jk}-\beta^j\tilde{\mc{K}}^{0k}-\beta^k\tilde{\mc{K}}^{0j}+\beta^j\beta^k\tilde{\mc{K}}^{00} \big)
=\alpha q \big( \tilde{H}^0+\sum_k w^k \tilde{H}^k  \big)\,.
\eea
Finally, we present the Maxwell's equations
\be \sum_i\fft{\partial}{\partial x^i}\left[\fft{h_1h_2h_3}{q q_i h_i}\,\Big((1+w^i\beta^i)\big(\tilde{B}^i-(\mathbf{w}\times\tilde{\mathbf{E}} )^i\big)-\bar{\beta}^i\, \mathbf{w}\cdot\tilde{\mathbf{B}} \Big) \right]=0 \,,\ee 
\be 
\fft{q}{h_1h_2h_3}\sum_i\fft{\partial}{\partial x^i}\left[\fft{h_1h_2h_3}{q q_i h_i}\,\Big((1+w^i\beta^i)\big(\tilde{E}^i-(\mathbf{w}\times\tilde{\mathbf{B}} )^i\big)-\bar{\beta}^i\, \mathbf{w}\cdot\tilde{\mathbf{E}} \Big) \right]=\tilde{J}^0+\mathbf{w}\cdot\tilde{\mathbf{J}}\,,
\label{chargegauss}\ee
\bea 
&&\alpha q\Big[(1+w^i\beta^i)\tilde{J}^i+\bar{\beta}^i\tilde{J}^0\Big]+\fft{\partial}{\partial t}\Big[(1+w^i\beta^i)\big(\tilde{E}^i-(\mathbf{w}\times\tilde{\mathbf{B}} )^i\big)-\bar{\beta}^i\, \mathbf{w}\cdot\tilde{\mathbf{E}} \Big]\\
&&=\fft{ qq_ih_i}{h_1h_2h_3}\sum_{jk}\fft{\partial}{\partial x^j}\left[\fft{\alpha h_1h_2h_3}{ q_i q_jh_ih_j}\, \Big(\epsilon^{ijk}(1+w^i\beta^i)(1+w^j\beta^j)\tilde{B}^k+(1+w^j\beta^j)\bar{\beta}^i\tilde{E}^j-(1+w^i\beta^i)\bar{\beta}^j\tilde{E}^i\Big)\right]\,,\nn
\label{curle}\eea
\bea 
&&\fft{\partial}{\partial t}\Big[(1+w^i\beta^i)\big(\tilde{B}^i-(\mathbf{w}\times\tilde{\mathbf{E}} )^i\big)-\bar{\beta}^i\, \mathbf{w}\cdot\tilde{\mathbf{B}} \Big] \\
&&=\fft{ qq_ih_i}{h_1h_2h_3}\sum_{jk}\fft{\partial}{\partial x^j}\left[\fft{\alpha h_1h_2h_3}{ q_i q_jh_ih_j}\, \Big(\epsilon^{ijk}(1+w^i\beta^i)(1+w^j\beta^j)\tilde{E}^k+(1+w^j\beta^j)\bar{\beta}^i\tilde{B}^j-(1+w^i\beta^i)\bar{\beta}^j\tilde{B}^i\Big)\right]\,.\nn
\label{curle}\eea
While these equations seem much involved, the geometric symmetry of the configuration considered in this work (see Fig. \ref{sheet} ) will greatly simplify the results for magnetic reconnection.

\section{Magnetic reconnection for a rotating reconnection layer}

\begin{figure}
\centering
\includegraphics[width=330pt]{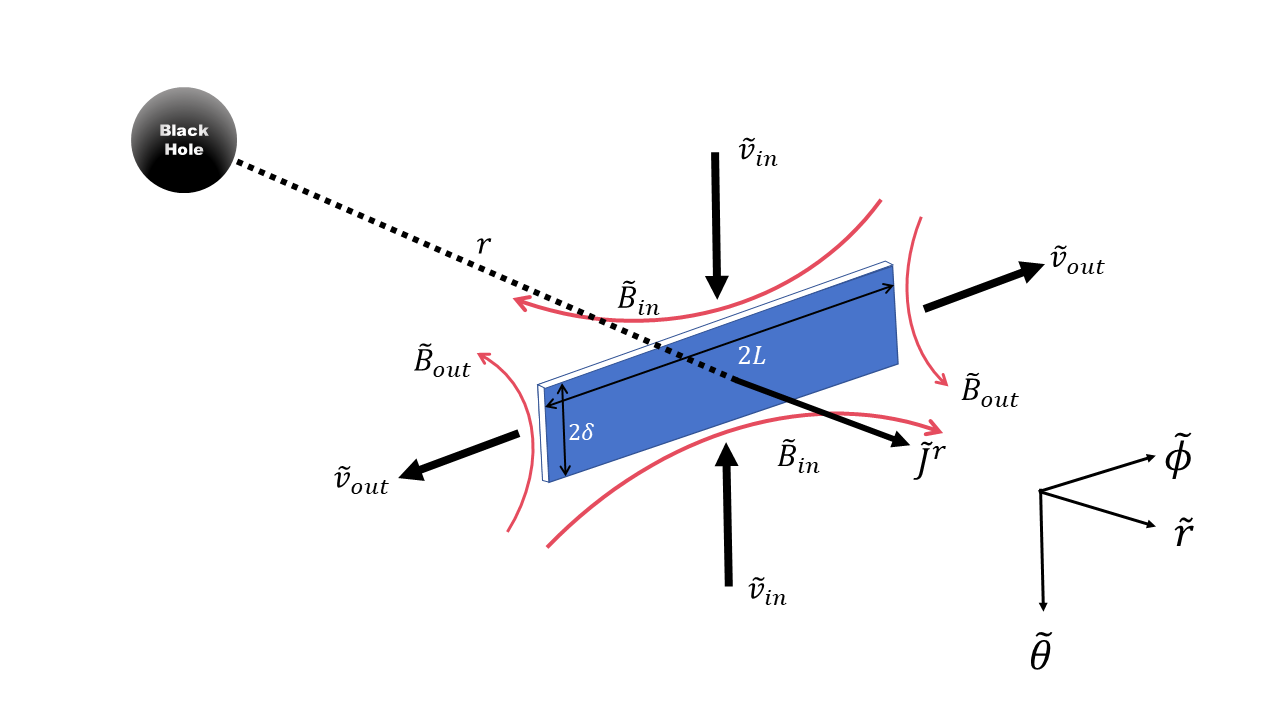}
\caption{The rotating reconnection layer in a Kerr black hole.}
\label{sheet}
\end{figure}

We are now ready to study the gravitational and the rotational effects in magnetic reconnection for a rotating reconnection layer. Consider the quasi-two-dimensional current sheet in the configuration depicted in Fig. \ref{sheet}. We work in quasi-stationarity $\partial_t\approx 0$ and assume $\tilde{v}^r=0=\tilde{B}^r$ and $\tilde\partial_{r}\approx 0$. For this two-dimensional configuration, the electric current density and the reconnected electric field exist only in the radial direction $\tilde{J}^i=\tilde{J}^r\delta^{ir}\,,\tilde{E}^i=\tilde{E}^r\delta^{ir}$. We set $\tilde{v}_{\mm{in}}=\tilde{v}_i \,e_\theta\,,\tilde{B}_{\mm{in}}=-\tilde{B}_0\, e_\phi$ and $\tilde{v}_{\mm{out}}=\tilde{v}_o \,e_\phi\,,\tilde{B}_{\mm{out}}=\tilde{B}_o\, e_\theta$, where $e_\theta\,,e_\phi$ are unit vectors in the corresponding direction. We will focus on the asymptotically weak gravity region in which $\beta\sim 2ar_g^2/r^2$. Without loss of generality, we consider a slowly rotating reconnection layer which obeys $w\sim\beta\ll \tilde{v}_o$. Under these conditions, we will derive the reconnection rate $\tilde{v}_i$ and various important properties of the outflow half-analytically.

Consider the continuity equation at first. The inflow flux at $i$ must balance the outflow flux at $O$. This gives
\be \delta\approx \fft{\tilde{\gamma}_i\tilde{v}_i}{\tilde{\gamma}_o\tilde{v}_o}\,q^2L \,,\label{thickness}\ee
where we have dropped a quadratic term $w\beta$ in the outflow flux since it is far less than unity in the asymptotic region. Here for notational simplicity, we have dropped the tilde for both the thickness and the length of the reconnection layer in the LCRF. Notice that the proper length in the LCRF is equal to $q^{-1}$ times the length in the BL frame , according to (\ref{crf}). While this relation is not written explicitly, it will be frequently adopted in our following discussions. It is worth emphasizing that the presence of $q^2$ factor in the relation (\ref{thickness}) originates from the fact that the spatial geometry is non-Euclidean for a comoving observer. We may interpret the result as the effective length of the current sheet is decreased $L_{\mm{eff}}\equiv q^2 L$ so that the reconnection rate will be increased because of $\tilde{v}_i\sim \delta/L_{\mm{eff}}$. Later we will show that this can indeed explain the leading contribution of the centrifugal electromotive force to the reconnection rate.

Likewise, magnetic flux conservation leads to
\be \tilde{B}_o\approx\fft{\delta}{L}\,\tilde{B}_0 \,,\label{magneticflux}\ee
where we have assumed $\partial_\theta (w \tilde{E}_r)\ll \partial_\theta \tilde{B}^\theta$, which holds in the asymptotically weak gravity region, in which $w\ll 1$. 

We assume the current density $\tilde{J}^\mu$ is nearly uniform so that from the Maxwell equation (\ref{curle})
\be \tilde{J}^r\Big|_X \approx \fft{\tilde{B}_0}{\delta} \,.\label{currentr}\ee
Notice that all the above results are exactly the same as those in the flat spacetime limit. This is partly due to symmetry of the configuration and partly due to the fact that we work in the weak gravity region.

To derive the outflow velocity, we consider the $\phi$-direction of the momentum equation (\ref{momenta}). It turns out that all terms involving the gravitational tidal force (associated to the metric derivatives) are vanishing identically due to symmetry of the configuration. The momentum equation greatly simplifies to 
\be \sum_j\fft{\partial}{\partial x^j}\left[\fft{\alpha h_1h_2h_3}{qq_jh_j}\big( \tilde{\mc{T}}^{\phi j}-\beta^\phi \tilde{\mc{P}}^j \big) \right] =0 \,.\ee
Since $w\sim \beta\ll \tilde{v}_o$, the solution to leading order at large $r$ is given by $\tilde{T}^{\phi\phi}\approx \mm{const}$ at the neutral line. This is a very good approximation since the outflow is mildly relativistic. Evaluating the stress tensor using (\ref{stresstensor}) and $h\approx 4p\approx 2\tilde{B}_0^2$ for hot relativistic plasmas, we can readily find that $\tilde{\gamma}_o\tilde{v}_o\approx 1/\sqrt{2}$. Again the weak gravity force has little impact on the outflow velocity. 

However, unlike these features, we will show that the reconnection rate is strongly influenced by the gravitation and rotation of the reconnection layer. Consider r-direction of the Ohm's law (\ref{spatialohm}). On the current sheet, it gives
\be\label{ergeneral} \tilde{E}_r=-\tilde{v}_\theta \tilde{B}_\phi+\tilde{v}_\phi \tilde{B}_\theta+\eta\tilde{J}^r+\fft{1}{ne\alpha h_1h_2h_3}\sum_j\fft{\partial}{\partial x^j}\Big(\fft{\alpha h_1h_2h_3}{q_j h_j}\,\tilde{\mc{K}}^{rj} \Big) +\fft{h\xi\tilde{J}^0}{2n^2e^2q^2\alpha h_1}\fft{\partial \alpha}{\partial r}+\cdots \,,\ee
where we have taken $\tilde{\gamma}\approx 1$. On the right hand side, the fourth term describes the transport of momentum by the current whereas the fifth term characterizes the gravitational electromotive force. The dots stands for all the terms involving $G_{ij}$ and $\sigma_{ij}$, which behave as $r_g^4 /r^5$ to leading order at large $r$. This is however far less than the gravitational force term $\sim r_g/r^2$ and hence is negligible. At the incidence point $i$, $\tilde{v}_\phi=0$ and $\eta=0$ since the plasma is ideal beyond the diffusion region. This gives
\be \tilde{E}_r\Big|_i=\tilde{v}_i\tilde{B}_0+\fft{h\xi r_g}{2n^2e^2 r^2}\,\tilde{J}^0\Big|_{i} \,.\label{eri}\ee
Similarly at the center of the diffusion region $X$, using $\tilde{v}_\theta=0=\tilde{v}_\phi$, we find
\be \tilde{E}_r\Big|_X=\big(\eta+q^{-2}\Lambda \big)\tilde{J}^r+\fft{h\xi r_g}{2n^2e^2 r^2}\,\tilde{J}^0\Big|_{X} \,,\label{erx}\ee
where the parameter $\Lambda$ is defined as
\be \Lambda=\fft{h\xi}{4n^2e^2L} \,.\ee
It characterizes the contribution of thermal inertia effect to the magnetic reconnection. It is interesting to notice that in (\ref{erx}), the $q^{-2}$ factor in front of $\Lambda$ follows from the effective length of the reconnection layer. We may introduce $\Lambda_{\mm{eff}}\equiv q^{-2}\Lambda$, which describes the enhanced thermal inertia effect.

Since there is no quasi-neutrality $\tilde{J}^0\neq 0$ in the black hole background, generally the electric fields are not equal $\tilde{E}_r|_i\neq \tilde{E}_r|_X$. The difference arises owing to different gravitational gradients at the two points because of $r_i-r\approx \delta^2/8r$, where $r_X=r$. The effect of the gravitational electromotive force can be extracted from the Gauss's law (\ref{chargegauss}). Integrating the equation along the radial direction, one finds
\be  \tilde{E}_r\Big|_i-\tilde{E}_r\Big|_X\approx \fft{h_1\delta^2 }{8r}\,\tilde{J}^0 \,.\label{erdiff}\ee
Finally, we have to connect the separation of the charge density $\tilde{J}^0$ to the current density $\tilde{J}^r$ using the temporal Ohm's law (\ref{timeohm}). We assume that compared to the gravitational gradient, the variation of the current density $\tilde{J}^r$ is negligible. Explicitly one has
\be \fft{q}{h_1h_2h_3}\fft{\partial}{\partial r}\Big[\fft{\alpha h_1h_2h_3}{qh_1}\,\tilde{\mc{K}}^{tr} \Big]\ll \fft{1}{h_1}\fft{\partial\alpha}{\partial r}\,\tilde{\mc{K}}^{tr} \,.\ee 
Using $\rho_e=-\tilde{U}_\mu\tilde{J}^\mu=\tilde{\gamma}\tilde{J}^0$, we deduce at the outflow point O
\be\fft{q^2}{neh_1h_2h_3}\fft{\partial}{\partial \phi}\Big[\fft{\alpha h_1h_2h_3}{q^2h_3}\,\tilde{\mc{K}}^{t\phi} \Big]\Big|_O+\fft{q\chi}{ne h_1}\fft{\partial \alpha}{\partial r}\,\tilde{\mc{K}}^{tr}\Big|_O\approx  \alpha q\eta\tilde{\gamma}^2\tilde{v}^2\tilde{J}^0\Big|_O \,,\label{timeohm1}\ee
where $\chi$ includes all contributions from $G_{ij}$ and $\sigma_{ij}$ terms in (\ref{timeohm}). However, these term simply give subleading order contributions in the weak gravity region because of $ \chi=1+12a^2r_g^3/r^3$ at large r. Using (\ref{timeohm1}), evaluation of the charge density at $X$ yields
\be \tilde{J}^0\Big|_X\approx \fft{2\Lambda L \chi_g}{\alpha h_1(\eta+q^{-2}\Lambda)}\fft{\partial \alpha}{\partial r}\,\tilde{J}^r \,,\label{jtjr}\ee
where $\chi_g\approx 1-L^2/4r-L^2r_g/8r^3 \,,$
where we have expanded the radial distance of the outflow point at the X point $r_o\approx r+L^2/8r$. Notice that a nonzero separation of the charge density arises from the gravitational tidal force. Finally, combining all these results and using (\ref{thickness}), we arrive at the reconnection rate 
\be \tilde{v}_i=\Big(\fft{1}{q^2S}+\fft{\Lambda}{q^4 L}\Big)^{1/2}\left[1+\fft{\Lambda L^2r_g}{8(\eta+\Lambda)r^3}+\cdots\right] \,,\label{finalrate}\ee
where $S=L/\eta$ is the relativistic Lindquist number. Notice that the first (square root) term in (\ref{finalrate}) is exact in large r. The square bracket encodes all contributions from the gravitational electromotive force, which increases the reconnection rate in comparison to the flat spacetime limit $r_g\rightarrow 0$. To leading order, the result is essentially the same as the Schwarzschild case \cite{Asenjo:2019nji}. It was interpreted as that the gravitational force acting on the inflow point and the X point is slightly different in $\theta$-direction so that a net attraction is produced along the plane of the reconnection layer \cite{Asenjo:2019nji}. Here our new finding is the black hole spin decreases the reconnection rate slightly but the effect is much weaker than the mass since it contributes to $1/r^6$ order at most ( in this term extra rotation of the configuration slightly deceases the reconnection rate as well but the effect is even weaker than the black hole spin as long as $w\sim\beta\sim r_g^2/r^2$). 


On the other hand, the leading contribution of the centrifugal electromotive force is given by the first term in (\ref{finalrate}), which increases the reconnection rate as well because of $q\leq 1$. In particular, in the flat spacetime limit, one has
\be \tilde{v}_i\Big|_{r_g\rightarrow 0}=\left(\fft{\eta}{L_{\mm{eff}}}+\fft{\Lambda_{\mm{eff}}}{ L_{\mm{eff}}}\right)^{1/2} \,.\ee
It implies that while the centrifugal electromotive force increases the reconnection rate as the gravitational force, the underlying mechanism is substantially different: it acts on the charge current density by decreasing the effective length of the current sheet owing to the non-Euclidean spatial geometry for the comoving observer. This results to twofold contributions associated to the charged carrier transport (collision) and the thermal-inertia effect (collisionless), respectively. In particular, unlike the gravitational force, these effects do not rely on the breaking of quasi-neutrality of the plasmas (separation of the charge density $\tilde{J}^0$ vanishes in the flat spacetime limit, as established in (\ref{jtjr})). We may define an effective electrical resistance $\eta_{\mm{eff}}\equiv \eta+\Lambda_{\mm{eff}}$ so that the reconnection rate behaves as $\tilde{v}_i\sim \delta/L_{\mm{eff}}\sim \big( \eta_{\mm{eff}}/L_{\mm{eff}}\big)^{1/2}$.

To end this section, consider slow rotation of the reconnection layer. One has in the flat spacetime limit
\be \tilde{v}_i\Big|_{r_g\rightarrow 0}= \Big(\fft{1}{S}+\fft{\Lambda}{L}\Big)^{1/2}\left[ 1+\fft{(L+2\Lambda S)}{2(L+\Lambda S)}\,w^2+\cdots\right]\,.\ee
It follows that the effect of the centrifugal electromotive force could be either stronger or weaker than the gravitational force, depending on the asymptotic behavior of the velocity $w$. In particular, for a critical choice $w\sim 1/r^{3/2}$, 
the leading order contribution will become the same as the gravitational electromotive force. It implies that the effect of gravitational force in magnetic reconnection for a unrotating reconnection layer could be partly captured by a rotating reconnection layer with a proper angular velocity in the laboratory frame.

\section{Conclusion}

We have examined various important properties of magnetic reconnection in a Kerr black hole background. We consider a generally rotating reconnection layer that does not corotate with the black hole. By focusing on the weak gravity region and a slowly rotating configuration, we extracted the leading contributions of the centrifugal and gravitational electromotive forces to the reconnection rate. While both increase the reconnection rate, the underlying mechanisms are substantially different. The gravitational force leads to the separation of charge density and breaks the quasi-neutrality of the plasma, whereas the centrifugal electromotive force acts on the electric current by decreasing the effective length of the current sheet, due to the non-Euclidean spatial geometry for a comoving observer. The latter enhances both the charged carrier transport and the thermal-inertia effect in the current sheet, irrespective of the existence of a black hole.

In this work, we focus on a reconnection layer that is far from the black hole. In this case, the contribution of the gravitational electromotive force heavily relies on the plasma model, which retains the thermal inertia effect in magnetic reconnection. It would be very interesting to relax this condition and study the strong gravitational effect. We have shown that for a circular stable orbit, the magnetic reconnection process is dramatically influenced by the rotation of the reconnection layer. Generalization of the results to the general motion of plasma (in the stationary approximation) will be very important. It is also interesting to connect the bulk physics in the locally corotating frame to observables (for example, the outflow energy) that are measured at asymptotic infinity. Following recent developments in this direction \cite{Comisso:2020ykg,Chen:2024ggq,Shen:2024sdr}, our results can be applied to study energy extraction from black holes via magnetic reconnection. We leave these to future research.


\section*{Acknowledgments}
The work is partly supported by NSFC Grant No. 12275004, 12205013 and 11873044. Z.Y. Fan was supported in part by the National Natural Science Foundations of China with Grant No. 11805041 and No. 11873025 and also supported in part by Guangzhou Science and Technology Project 2023A03J0016. MG is also endorsed by ``the Fundamental Research Funds for the Central Universities” with Grant No. 2021NTST13.

\appendix

\end{document}